\documentclass{aastex62}

\submitjournal{ApJ}

\shorttitle{Sample article}
\shortauthors{Zhang et al.}
\begin{document}

\title{Nearly 30,000 late-type main-sequence stars with stellar age from LAMOST DR5}

\correspondingauthor{Jingkun Zhao}
\email{zjk@bao.ac.cn}

\author[0000-0002-0786-7307]{Jiajun Zhang}
\affil{CAS Key Laboratory of Optical Astronomy, National Astronomical Observatories, Chinese Academy of Sciences, Beijing 100101, China. zjk@nao.cas.cn}
\affil{School of Astronomy and Space Science, University of Chinese Academy of Sciences, Beijing 100049, China}

\author{Jingkun Zhao}
\affiliation{CAS Key Laboratory of Optical Astronomy, National Astronomical Observatories, Chinese Academy of Sciences, Beijing 100101, China. zjk@nao.cas.cn}

\author{Terry D. Oswalt}
\affiliation{Embry-Riddle Aeronautical University, Aerospace Boulevard, Daytona Beach FL, USA, 32114. oswaltt1@erau.edu}

\author{Xilong Liang}
\affil{CAS Key Laboratory of Optical Astronomy, National Astronomical Observatories, Chinese Academy of Sciences, Beijing 100101, China. zjk@nao.cas.cn}
\affil{School of Astronomy and Space Science, University of Chinese Academy of Sciences, Beijing 100049, China}

\author{Xianhao Ye}
\affil{CAS Key Laboratory of Optical Astronomy, National Astronomical Observatories, Chinese Academy of Sciences, Beijing 100101, China. zjk@nao.cas.cn}
\affil{School of Astronomy and Space Science, University of Chinese Academy of Sciences, Beijing 100049, China}

\author{Gang Zhao}
\affil{CAS Key Laboratory of Optical Astronomy, National Astronomical Observatories, Chinese Academy of Sciences, Beijing 100101, China. zjk@nao.cas.cn}
\affil{School of Astronomy and Space Science, University of Chinese Academy of Sciences, Beijing 100049, China}

%% Note that the \and command from previous versions of AASTeX is now
%% depreciated in this version as it is no longer necessary. AASTeX
%% automatically takes care of all commas and "and"s between authors names.

%% AASTeX 6.2 has the new \collaboration and \nocollaboration commands to
%% provide the collaboration status of a group of authors. These commands
%% can be used either before or after the list of corresponding authors. The
%% argument for \collaboration is the collaboration identifier. Authors are
%% encouraged to surround collaboration identifiers with ()s. The
%% \nocollaboration command takes no argument and exists to indicate that
%% the nearby authors are not part of surrounding collaborations.

%% Mark off the abstract in the ``abstract'' environment.
\begin{abstract}
We construct a sample of nearly 30,000 main-sequence stars with 4500K $<T\rm_{eff}<$ 5000K and stellar ages estimated by the chromospheric activity$-$age relation. This sample is used to determine the age distribution in the $R-Z$ plane of the Galaxy, where $R$ is the projected Galactocentric distance in the disk midplane and $Z$ is the height above the disk midplane. As $|Z|$ increases, the percentage of old stars becomes larger. It is known that scale-height of Galactic disk increases as $R$ increases, which is called flare. A mild flare from $R$ $\sim$ 8.0 to 9.0 kpc in stellar age distribution is found. We also find that the velocity dispersion increases with age as confirmed by previous studies. Finally we present spiral-shaped structures in $Z-\upsilon_{Z}$ phase space in three stellar age bins. The spiral is clearly seen in the age bin of [0, 1] Gyr, which suggests that a vertical perturbation to the disk probably took place within the last $\sim$ 1.0 Gyr.
\end{abstract}

%% Keywords should appear after the \end{abstract} command.
%% See the online documentation for the full list of available subject
%% keywords and the rules for their use.

%% From the front matter, we move on to the body of the paper.
%% Sections are demarcated by \section and \subsection, respectively.
%% Observe the use of the LaTeX \label
%% command after the \subsection to give a symbolic KEY to the
%% subsection for cross-referencing in a \ref command.
%% You can use LaTeX's \ref and \label commands to keep track of
%% cross-references to sections, equations, tables, and figures.
%% That way, if you change the order of any elements, LaTeX will
%% automatically renumber them.
%%
%% We recommend that authors also use the natbib \citep
%% and \citet commands to identify citations.  The citations are
%% tied to the reference list via symbolic KEYs. The KEY corresponds
%% to the KEY in the \bibitem in the reference list below.

\section{Introduction} \label{sec:intro}

Stellar ages are very important to understanding the formation and evolution of the Milky Way \citep{Soderblom2010}. Evolutionary isochrones are frequently used to determine age. By placing a star on model isochrones in the Hertzsprung-Russell Diagram (HRD), the age or a limit on age can be estimated by interpolation. This method is well-suited for older stars, massive stars and evolved stars. However, for young stars (ages less than $\sim$ 1/3 of their overall main-sequence (MS) lifetime) and low-mass stars, it is unsuitable \citep{Soderblom2010}.

For late-type MS stars, several empirical relations are usually used to derive ages: gyrochronology, chromospheric activity (CA) and lithium depletion \citep{Skumanich1972,Barnes2007,Mamajek2008,Soderblom2010}. Gyrochronology is a technique based on the relation between stellar rotation period and age \citep{Barnes2007}. Stellar rotation slows due to magnetic braking as a star grows older. \cite{Barnes2007} obtained an empirical relation between rotation period, $B-V$ color and age using open clusters and the sun whose ages are well-known. As a star grows older, the CA declines, so it can serve as an age indicator \citep{Skumanich1972,Soderblom1991,Mamajek2008}. Combining the cluster CA data with modern cluster age estimates, \cite{Mamajek2008} derived a CA$-$age relation for F7-K2 dwarfs with typical precision of $\sim$ 0.2 dex between $\sim$ 0.6-4.5 Gyr. They used the CA$-$age relation to estimate ages for 108 solar-type field dwarfs within 16 pc. The number of stars was small because of color (0.5 $<B-V<$ 0.9), parallax, and absolute magnitude constraints.

The Large Sky Area Multi-Object Fiber Spectroscopic Telescope Data Release 5 (LAMOST DR5) has obtained more than eight million stellar spectra with a resolution of $\sim$ 1800 at the 5500\AA\ \citep{zhao2006,zhao2012,cui2012,Deng2012,Luo2015}. These spectra cover the wavelengths of Ca \uppercase\expandafter{\romannumeral2} HK lines (3968\AA\ and 3934\AA) and are sufficient to measure CA \citep{Fang2018,Zhang2019,Zhangjinghua2020}. \cite{Zhang2019} (hereafter Paper I) calculated a CA index, $\log R'\rm_{CaK}$, for member stars of open clusters and studied the CA$-$age relations in different effective temperature ($T\rm_{eff}$) ranges. This provides a way to roughly estimate ages for late-type MS stars in the LAMOST survey. Although the derived ages have relatively large uncertainties and could be affected by factors such as mass, binary companions, etc., it is possible to estimate ages for a large number of field stars for statistical analysis.

Many authors have studied the relation between stellar kinematics and age using red giant stars, sub-giant stars or MS turn-off stars instead of late-type MS stars, because ages are easier to estimate and they can be observed to larger distances than late-type MS stars \citep{Xiang2017,Tian2018,Yu2018MNRAS,Mackereth2019,Wu2019}. For example, \cite{Xiang2017} estimated ages for over 900,000 MS turn-off and sub-giant stars from the LAMOST Survey. They found stellar ages exhibited positive vertical and negative radial gradients across the Galactic disk.  The age$-$velocity dispersion relation (AVR) has been studied by many authors \citep{Wielen1977,Nordstrom2004,Mackereth2019}. It is well known that velocity dispersion increases with age.

Vertical phase mixing is an important phenomena in the Galactic disk \citep{Antoja2018,Tian2018,Li2020}. It usually manifests as spiral structures of Galactocentric azimuthal velocities ($\upsilon_{\phi}$) or radial velocities ($\upsilon_{R}$) in the phase space of $Z-\upsilon_{Z}$, where $Z$ is the height above the disk midplane and $\upsilon_{Z}$ is the velocity component perpendicular to the midplane. \cite{Tian2018} found that the spiral structures existed in populations of stars with age less than 0.5 Gyr, which supported the conclusion that the vertical perturbation took place no later than 0.5 Gyr.

Few studies have used late-type MS stars to investigate the relation between kinematics and age. \cite{Zhao2013AJ}, using CA index $S_{\rm HK}$ measurements for over 13,000 F, G, and K disk stars in the Sloan Digital Sky Survey (SDSS) Data Release 7 (DR7) spectroscopic sample, found the fraction of K dwarfs that had strong CA dropped with vertical distance from the Galactic midplane. Those active stars were relative young. Because of dynamical heating in the thin disk, older stars are more likely to be found farther from the Galactic plane. We used late-type MS stars with estimated ages to study kinematics. The results were then compared to other studies which used sub-giant stars or red giant stars.

This paper is organized as follows: Section \ref{sec:sample} introduces the sample and how ages for late-type MS stars were estimated. The relation between stellar kinematics and age is discussed in section \ref{sec:Kinematics}. Our conclusions are summarized in section \ref{sec:conclusion}.

\begin{figure}[!t]
\center
\includegraphics[scale=0.5]{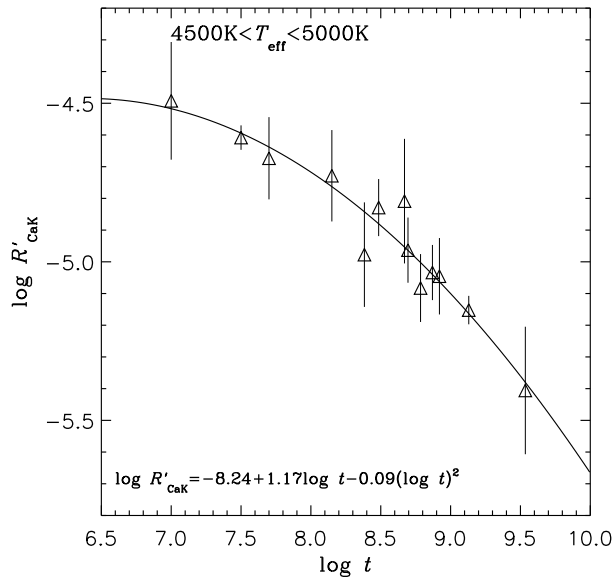}
\caption{Relation between CA index, $\log R'\rm_{CaK}$, and age ($\log t$) among open clusters (Details can be found in Paper I). Each triangle and error bar stands for mean value and standard deviation of $\log R'\rm_{CaK}$, respectively, for member stars within each cluster. The abscissa is age in log form. The Figure includes stars only with 4500K $<T\rm_{eff}<$ 5000K. The quadratic function fitted to these points is listed in the bottom-left corner. The solid line is the corresponding curve. \label{fig:logt_logr_CaK4500K_5000K}}
\end{figure}

%\newpage
\section{Late-type MS stars with estimated ages} \label{sec:sample}
\subsection{CA$-$age relation of stars with 4500K $<T\rm_{eff}<$ 5000K and [Fe/H] $>$ -0.2}

In Paper I, we calculated the CA index, $\log R'\rm_{CaK}$, based on the Ca \uppercase\expandafter{\romannumeral2} K line for member stars of 82 open clusters. These stars are all FGK dwarfs. In Paper I, the CA$-$age relation was investigated in different $T\rm_{eff}$ ranges. Figure \ref{fig:logt_logr_CaK4500K_5000K} displays relation between the CA index ($\log R'\rm_{CaK}$) and age ($\log t$) for stars with 4500K $<T\rm_{eff}<$ 5000K ($T\rm_{eff}$ was provided by LAMOST DR5). The updated member star catalogue provided by \cite{Cantat2020} was used. For M67, we used member stars provided by \cite{Carrera2019} instead of \cite{Cantat2020} because catalogue of former study included more member stars. Those stars labeled as `Flare*', `pMS*', `YSO', `RSCVn', `SB*', `EB*WUMa', `EB*', `EB*Algol' and `EB*betLyr' by Simbad were excluded. In Figure \ref{fig:logt_logr_CaK4500K_5000K}, each triangle and error bar stands for the mean value and standard deviation of $\log R'\rm_{CaK}$, respectively, for member stars within each open cluster, while the abscissa is the age ($\log t$). Table \ref{clusters} lists the relevant data for these open clusters: cluster name, logarithm of age (from HRD), age in million years, mean value of metallicity from LAMOST DR5, standard deviation in metallicity, number of stars, $\log R'\rm_{CaK}$ index, standard deviation in $\log R'\rm_{CaK}$, estimated age from CA and relative difference between estimated age and accurate age. We used a quadratic function to fit these data points, shown as a solid line in Figure \ref{fig:logt_logr_CaK4500K_5000K} and given in Equation \ref{Equation:CA_age_relation}, which was used to estimate ages for late-type MS stars with 4500K $<T\rm_{eff}<$ 5000K and [Fe/H] $>$ -0.2. This metallicity limit was chosen because open clusters have average [Fe/H] $>$ -0.2 (see Table \ref{clusters}). We then estimated ages for open clusters using Equation \ref{Equation:CA_age_relation} and the results were shown in Table \ref{clusters}. The largest deviations from the curve in Figure \ref{fig:logt_logr_CaK4500K_5000K} occurred in NGC\_1039, Roslund\_6 and NGC\_2281 (see Table \ref{clusters}). Note that only two clusters (NGC\_752 and M67) are larger than 1.0 Gyr. Open clusters older than 1 Gyr are more difficult to observe than young clusters.

\begin{equation}\label{Equation:CA_age_relation}
\log R'_{\rm CaK}=-8.24+1.17\log t-0.09(\log t)^2,\    4500\rm{K}<\textit{T}\rm_{eff}<5000\rm{K}\ and\    [Fe/H]>-0.2
\end{equation}

\begin{deluxetable}{cccccccccc}[!ht]
\tablecaption{open clusters of Figure \ref{fig:logt_logr_CaK4500K_5000K} \label{clusters}}
\tablehead{
\colhead{Name} & \colhead{$\log t$} & \colhead{$t$(Myr)}  & \colhead{$\rm [Fe/H]_{LA}$}  & \colhead{$\rm [Fe/H]\_std_{LA}$}  & \colhead{N}  & \colhead{$\log R'\rm_{CaK}$}  & \colhead{$\log R'\rm_{CaK}\_std$}  & \colhead{$t'$(Myr)}  & \colhead{Relative difference(\%)}
}
\startdata
  ASCC\_16 & 7.0 & 10 & -0.13 & 0.074 & 4 & -4.49 & 0.186 & 5 & -46\\
  ASCC\_19 & 7.5 & 32 & -0.12 & 0.025 & 3 & -4.61 & 0.039 & 38 & +20\\
  $\alpha$ Per & 7.7 & 50 & -0.03 & 0.062 & 13 & -4.67 & 0.130 & 71 & +42\\
  Pleiades & 8.15 & 141 & -0.08 & 0.066 & 28 & -4.73 & 0.144 & 112 & -20\\
  NGC\_1039 & 8.383 & 242 & -0.03 & 0.096 & 12 & -4.98 & 0.165 & 544 & +125\\
  ASCC\_23 & 8.485 & 305 & -0.09 & 0.080 & 2 & -4.83 & 0.090 & 226 & -25\\
  Roslund\_6 & 8.67 & 468 & -0.05 & 0.047 & 5 & -4.81 & 0.196 & 198 & -57\\
  NGC\_1662 & 8.695 & 490 & -0.04 & 0.003 & 2 & -4.96 & 0.103 & 503 & +1\\
  NGC\_2281 & 8.785 & 603 & -0.05 & 0.058 & 4 & -5.08 & 0.107 & 936 & +53\\
  Hyades & 8.87 & 741 & 0.07 & 0.119 & 3 & -5.03 & 0.087 & 732 & -1\\
  Praesepe & 8.92 & 832 & 0.13 & 0.064 & 17 & -5.05 & 0.121 & 777 & -6\\
  NGC\_752 & 9.13 & 1349 & -0.04 & 0.064 & 5 & -5.15 & 0.045 & 1306 & -3\\
  M67 & 9.535 & 3388 & -0.02 & 0.059 & 7 & -5.41 & 0.201 & 3866 & +12\\
\enddata
\tablecomments{First column is the name of the open cluster.
The $\log t$ and $t$ are age of the cluster in logarithmic and decimal format, respectively. The ages of all clusters except for Hyades come from \cite{Kha2013}, while the age of Hyades comes from \cite{Gos2018}.
$\rm [Fe/H]_{LA}$ and $\rm [Fe/H]\_std_{LA}$ are the mean value and standard deviation of [Fe/H] calculated from the LAMOST DR5. The subscript ``LA'' means LAMOST DR5.
N is the number of cluster member stars with 4500K $<T\rm_{eff}<$ 5000K within each cluster.
The $\log R'\rm_{CaK}$ and $\log R'\rm_{CaK}\_std$ are the mean value and standard deviation of $\log R'\rm_{CaK}$ of each cluster, respectively.
The $t'$ is the estimated age from Equation \ref{Equation:CA_age_relation}.
The final column is the relative difference between estimated age ($t'$) and accurate age ($t$) and calculated via $(t'-t)/t$. }
\end{deluxetable}

The error bars in Figure \ref{fig:logt_logr_CaK4500K_5000K} consist of two parts. One is measurement error, the other is the inherent variance of the CA index. Stars with the same ages, masses and metallicities do not necessarily have the same CA index due to different rotation rates. Because of this, an age estimate for a single star may have a large error. Here we make use of a very large sample to obtain some generalized results.

The reasons why we choose this $T\rm_{eff}$ range are as follows. First, at low $T\rm_{eff}$ the variation in the mean $\log R'\rm_{CaK}$ with $\log t$ is larger compared to that at higher $T\rm_{eff}$ range. Also, there is a weak correlation between $\log R'\rm_{CaK}$ and $T\rm_{eff}$, because the $R'\rm_{CaK}$ value contains a contribution from the continum, which depends on $T\rm_{eff}$. Finally, the oldest open cluster, M67, has seven member stars at 4500K $<T\rm_{eff}<$ 5000K but no well observed member stars at lower $T\rm_{eff}$. Narrowing the $T\rm_{eff}$ range to [4500, 5000] K makes the CA$-$age relation useful to about 4.0 Gyr.

\subsection{Estimating ages for late-type MS stars}\label{Estimating_ages}

By using the same method outlined in Paper I, we first calculated the CA index, $\log R'\rm_{CaK}$, for a large number of field stars from LAMOST DR5. These field stars are the same as those in Paper I to determine the basal lines (see Section 3.2 of Paper I). The parameters of these stars are 4000K $<T\rm_{eff}<$ 7000K, $\log g$ (surface gravity) $>$ 4.0, -0.8 $<$ [Fe/H] $<$ 0.5 and S/N g (signal to noise ratio in g band) $>$ 30, provided by LAMOST DR5. The sample included 1,599,649 stars in total.

Equation \ref{Equation:CA_age_relation} only applies to stars with 4500K $<T\rm_{eff}<$ 5000K and [Fe/H] $>$ -0.2. And we applied these limits to the original sample. Next, we selected only stars with -4.51 $>\log R'\rm_{CaK}>$ -5.41, which corresponds to ages 0.01 $<t<$ 4.0 Gyr. Equation \ref{Equation:CA_age_relation} was then used to estimate ages for resulting sample of 30,619 stars.

\begin{figure}[!t]
\gridline{\fig{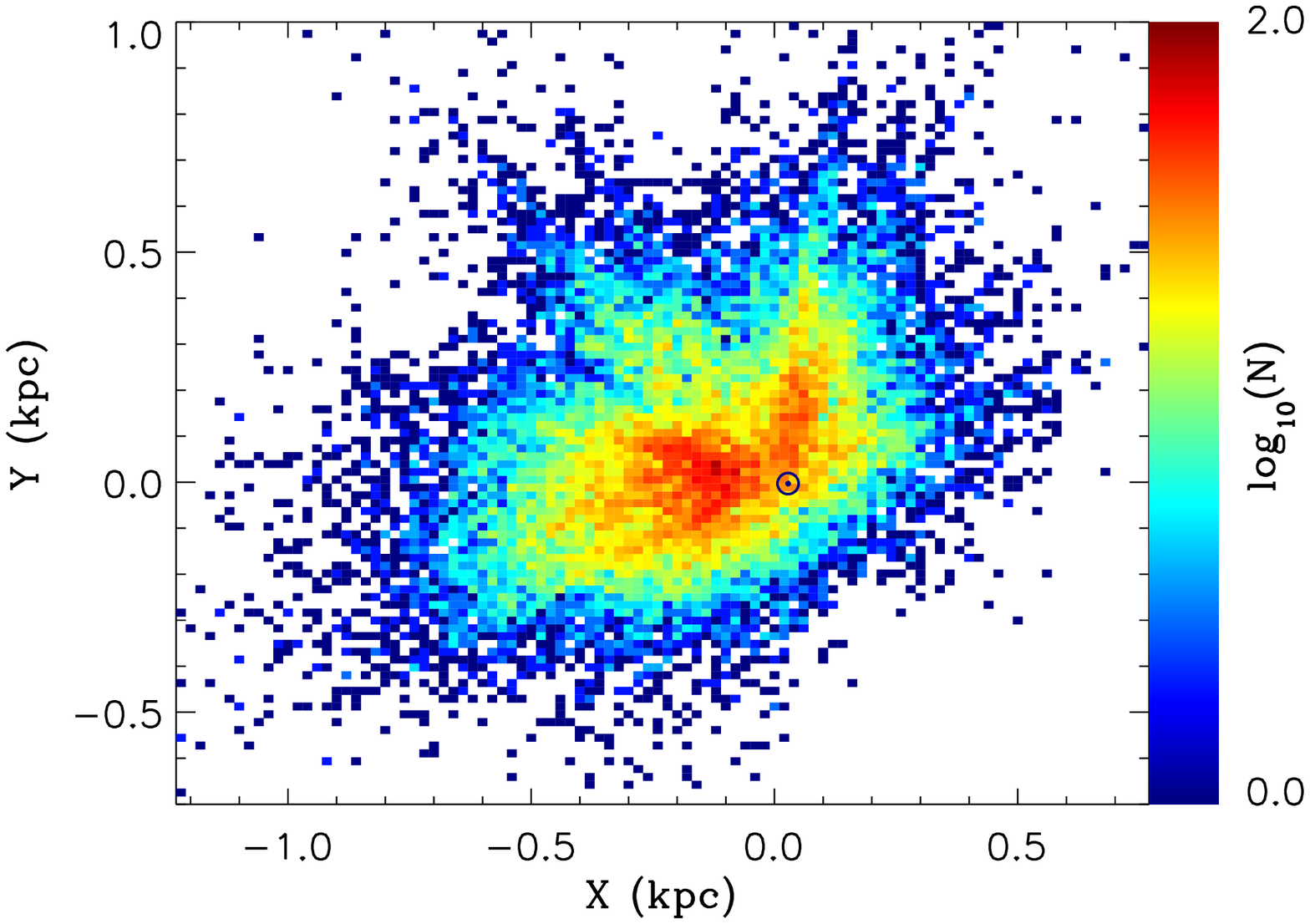}{0.5\textwidth}{(a)}
          \fig{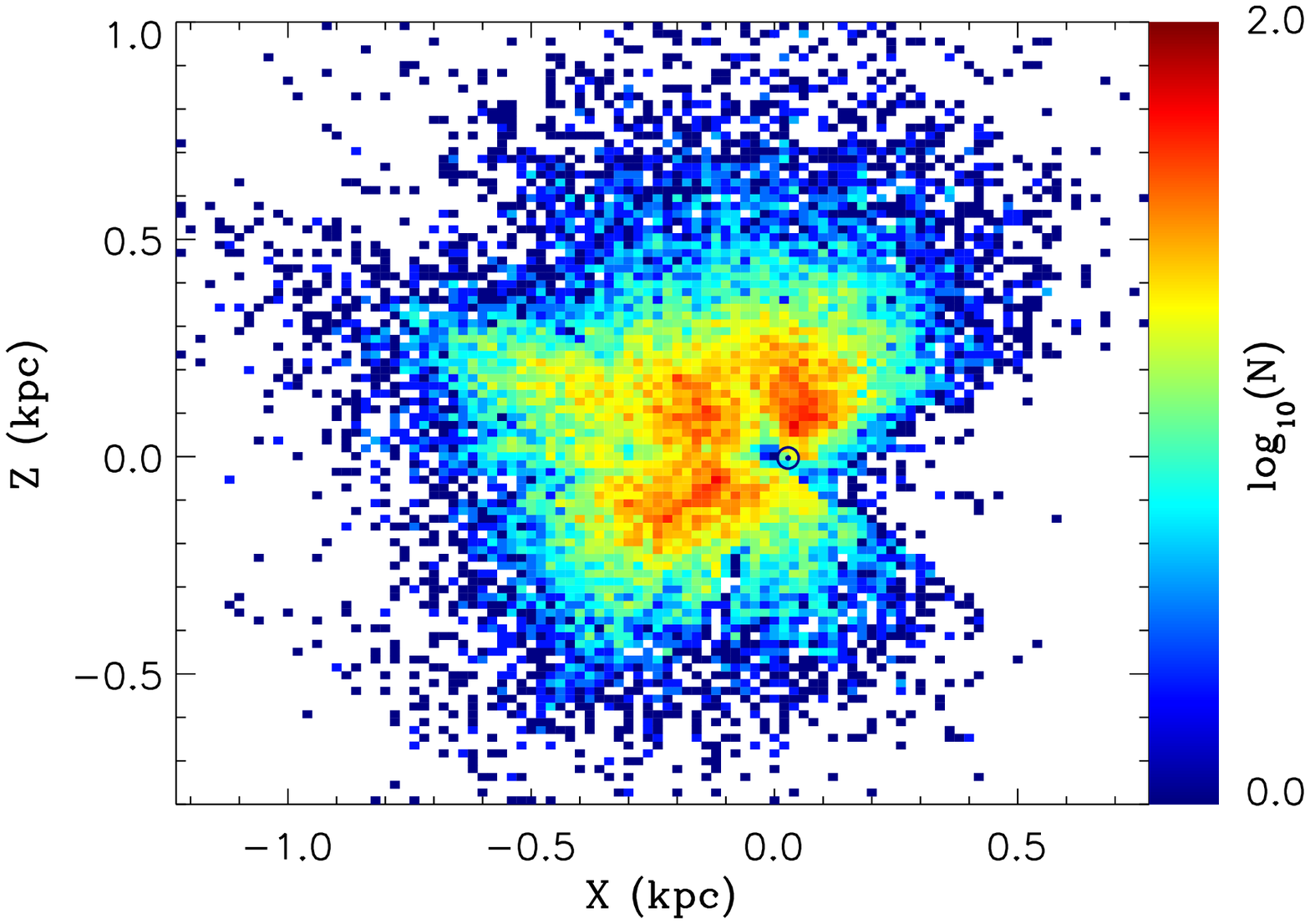}{0.5\textwidth}{(b)}
          }
\caption{Colour-coded stellar spatial-density distributions in the Galactic $X-Y$ and $X-Z$ planes for late-type MS stars observed with LAMOST DR5. The dotted circle marks the location of the sun at $X=Y=Z=0$. \label{fig:x_y_distribution}}
\end{figure}

Most of these selected stars have parallaxes and proper motions provided by \emph{Gaia} DR2 \citep{GaiaCollaboration2018}. We chose those with relative errors in parallax less than 20\% and calculated heliocentric positions ($X$, $Y$ and $Z$) relative to the sun. The $X$-axis points towards the Galactic center, the $Y$-axis points towards the Galactic rotation direction and the $Z$-axis points towards the Galactic north. These data are plotted in Figure \ref{fig:x_y_distribution}. The LAMOST survey mainly focuses on stars in the direction of the Galactic anti-center, so there are fewer stars towards the Galactic center ($+X$) as shown in Figure \ref{fig:x_y_distribution} \citep{Deng2012}. In Figure \ref{fig:x_y_distribution}(a), we see gaps in which there are fewer stars. For example, the two obvious gaps locate around (-0.5, 0.3) kpc and (-0.1, 0.5) kpc respectively. These gaps are caused by selection effect. The LAMOST observed stars of $-10^{\circ}<\delta<60^{\circ}$, which excluded the region near the north celestial pole \citep{zhao2012}. In northern summers due to weather and maintaining telescope, the observing time decreased \citep{zhao2012}. In the following analysis, we adopted the solar Galactocentric radius and vertical positions as ($R_{\sun}$, $Z_{\sun}$)=(8.27, 0.02) kpc \citep{Schonrich2010, Schonrich2012}. Using radial velocities provided by LAMOST DR5, velocities ($\upsilon_{R}$, $\upsilon_{\phi}$ and $\upsilon_{Z}$) relative to the Galactic center were calculated. We then applied the solar motion relative to the local standard of rest (LSR), where ($\rm U_{\sun}$, $\rm V_{\sun}$, $\rm W_{\sun}$)=(8.5, 13.38, 6.49) km/s \citep{Coskunoglu2011}. The $\upsilon_{R}$ is positive towards the Galactic anti-center, $\upsilon_{\phi}$ is positive towards the Galactic rotation direction and $\upsilon_{Z}$ is positive towards the Galactic north. In total, we compiled a sample of 29,041 stars with estimated ages, space positions ($X$, $Y$ and $Z$) and velocities ($\upsilon_{R}$, $\upsilon_{\phi}$ and $\upsilon_{Z}$).

\begin{figure}[!t]
\center
\includegraphics[scale=0.7]{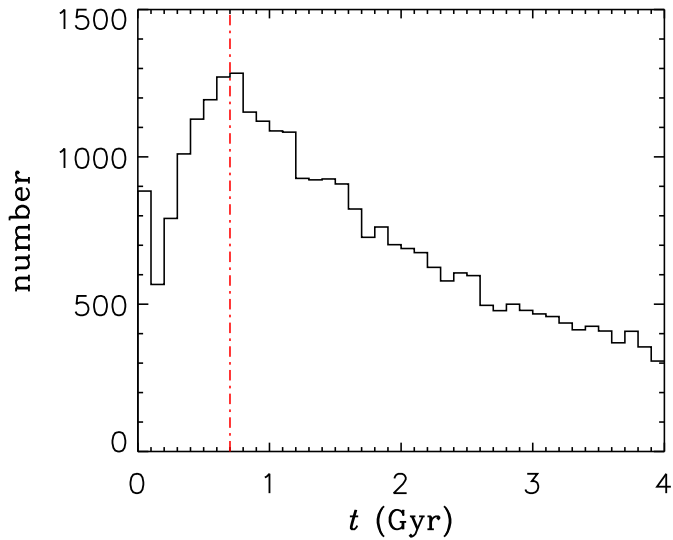}
\caption{Age distribution of stars in our sample. The bin size is 0.1 Gyr. Dash-dot line marks the location of $t=0.7$ Gyr. \label{fig:age_distribution}}
\end{figure}

The age distribution of this sample is shown in Figure \ref{fig:age_distribution}. There are two peaks in this distribution: one around $t=0.7$ Gyr and the other is the youngest bin ($t<0.1$ Gyr). Recent studies have suggested a recent star formation burst occurred about 0.6 Gyr ago \citep{Torres2016,Davenport2018}. \cite{Ruiz2020NatAs} found three conspicuous star formation bursts occurring at 5.7, 1.9 and 1.0 Gyr ago. The star formation burst occurring at 1.0 Gyr ago is consistent with the peak of $t=0.7$ Gyr in our age distribution within the formal uncertainty. They hypothesized that the Sagittarius dwarf galaxy pericentre passages induced the three star formation bursts. They also found a fourth possible star formation burst within the last 70 Myr, which may correspond to the peak at $t<0.1$ Gyr in our age distribution.

\section{Kinematics and Age} \label{sec:Kinematics}
\subsection{Age distribution in the $R-Z$ plane}

\begin{figure}[!ht]
\gridline{\fig{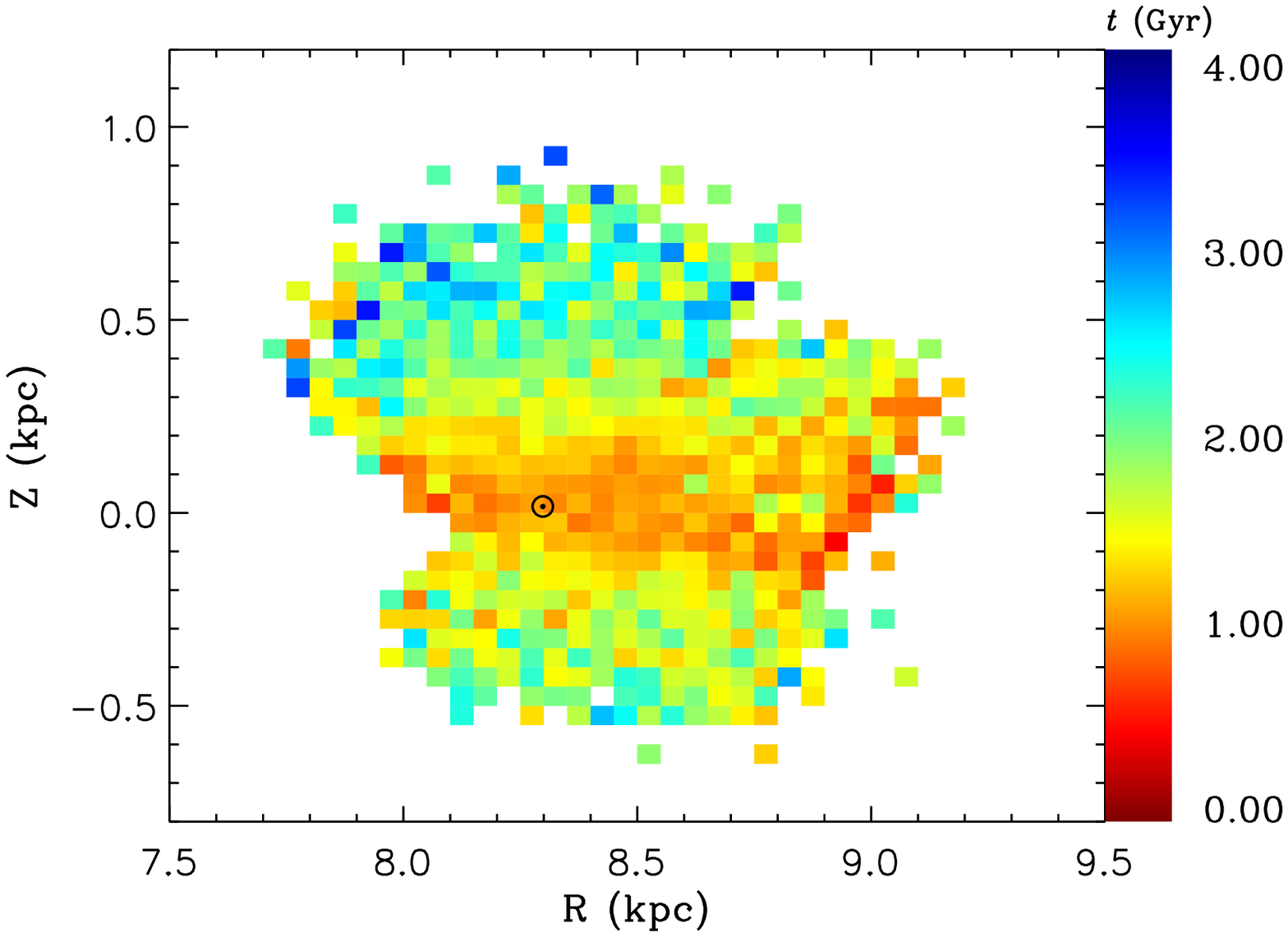}{0.5\textwidth}{(a)}
          \fig{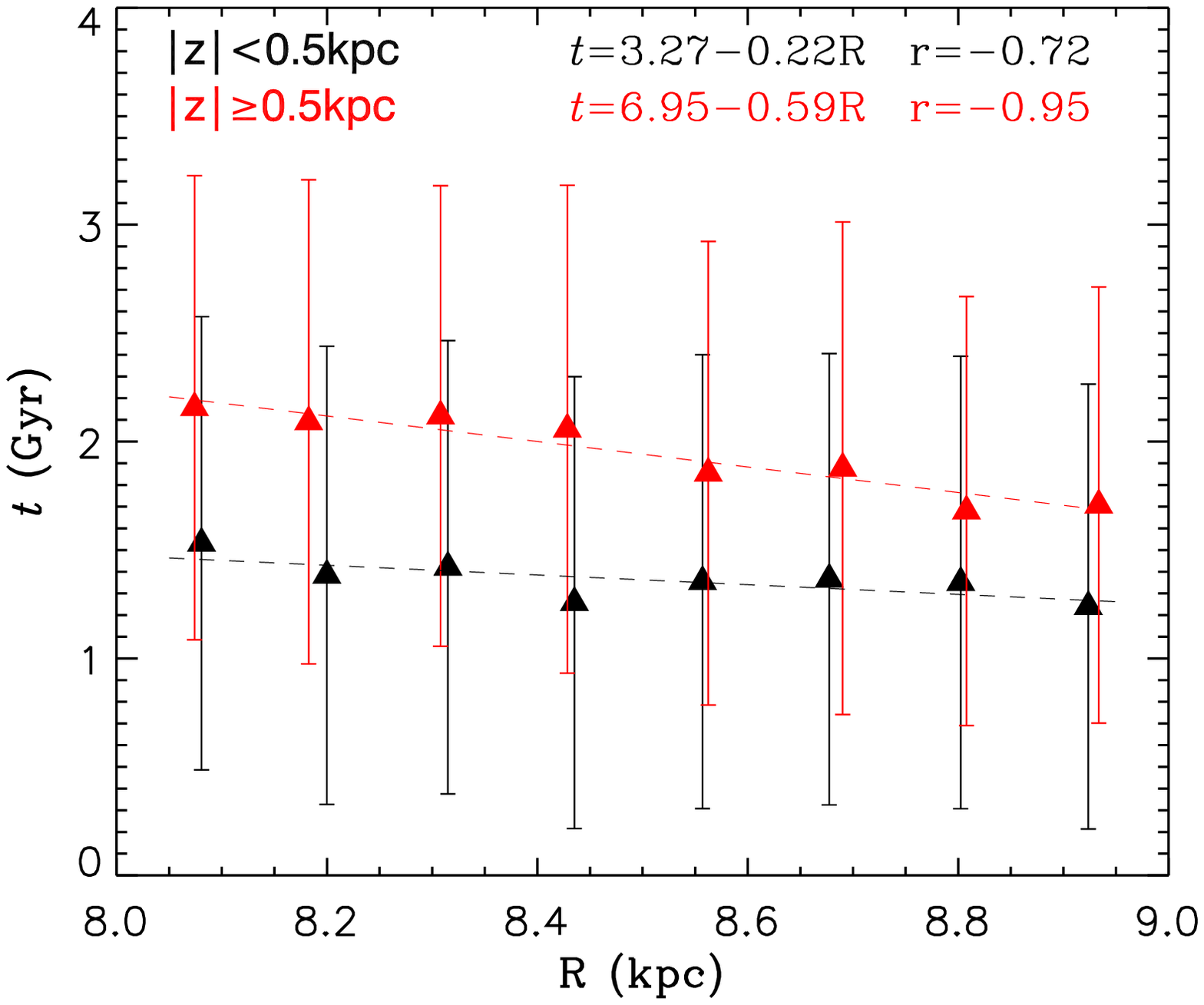}{0.5\textwidth}{(b)}
          }
\caption{(a): Colour-coded distribution of the median ages of stars in the $R-Z$ plane. The adopted bin size is 0.05 kpc in both the $R$-direction and the $Z$-direction. The number of stars in each bin is larger than five. The dotted circle marks the location of the sun. (b): Median age distribution as a function of $R$ for two sub-samples. The black triangles represent the sub-sample with $|Z|<$ 0.5 kpc, while the red triangles represent the sub-sample with $|Z|\geq$ 0.5 kpc. In both sub-samples, eight equal bins in [8.0, 9.0] kpc are shown. The error bars represent the standard deviations of each bin's age. Linear fits for both sub-samples are indicated by the two dashed lines. The fitting relations and Pearson correlation coefficients are listed at the top. \label{fig:R_Z_bin}}
\end{figure}

Figure \ref{fig:R_Z_bin}(a) displays the age distribution of our sample in the $R-Z$ plane. Here $R$ is the projected Galactocentric distance in the disk midplane, and $Z$ is the height above the disk midplane. Although the spatial distribution of our sample is small, we agree with the results of \cite{Xiang2017}. In the vertical direction, the percentage of old stars increases as $|Z|$ increases. Thin-disk dynamical heating is believed to cause this phenomena \citep{West2008,Casagrande2016}. It is known that scale-height of Galactic disk increases as $R$ increases, which is called flare \citep{Lopez2002}. Flare is found not only via star counts, but also in stellar age distribution \citep{Xiang2017}. We can find a mild flare from $R$ $\sim$ 8.0 to 9.0 kpc in Figure \ref{fig:R_Z_bin}(a). In order to further verify this flaring-age structure, we plot the median age distribution as a function of $R$ in Figure \ref{fig:R_Z_bin}(b). The sample was divided into two sub-samples with $|Z|<$ 0.5 kpc and $|Z|\geq$ 0.5 kpc. The expected trend that the median age decreases mildly with $R$ can be seen, which is a characteristic of the flaring-age structure \citep{Xiang2017,Wu2019}.

\subsection{Age$-$velocity dispersion relation}
\begin{figure}[!t]
\gridline{\fig{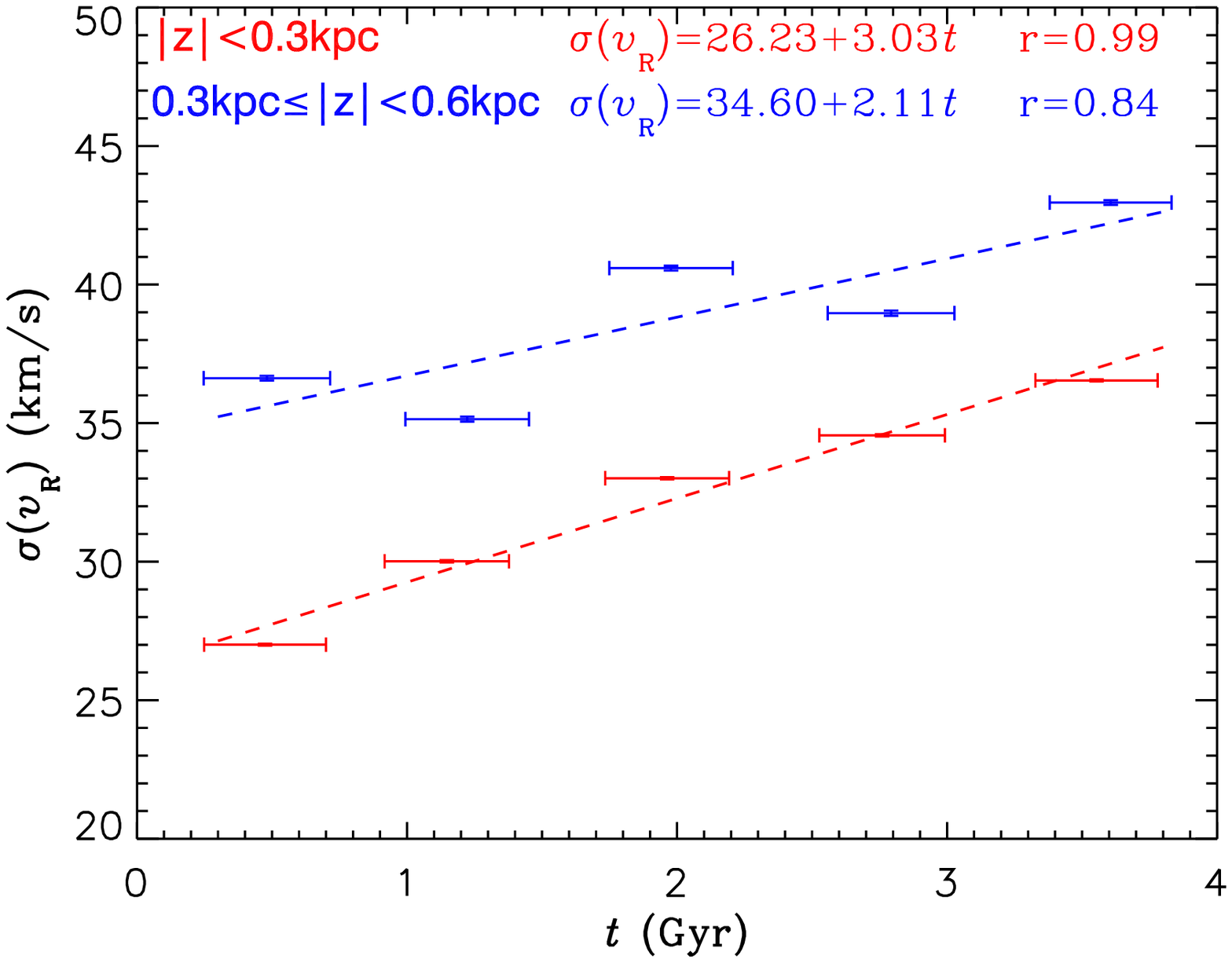}{0.5\textwidth}{(a)}
          \fig{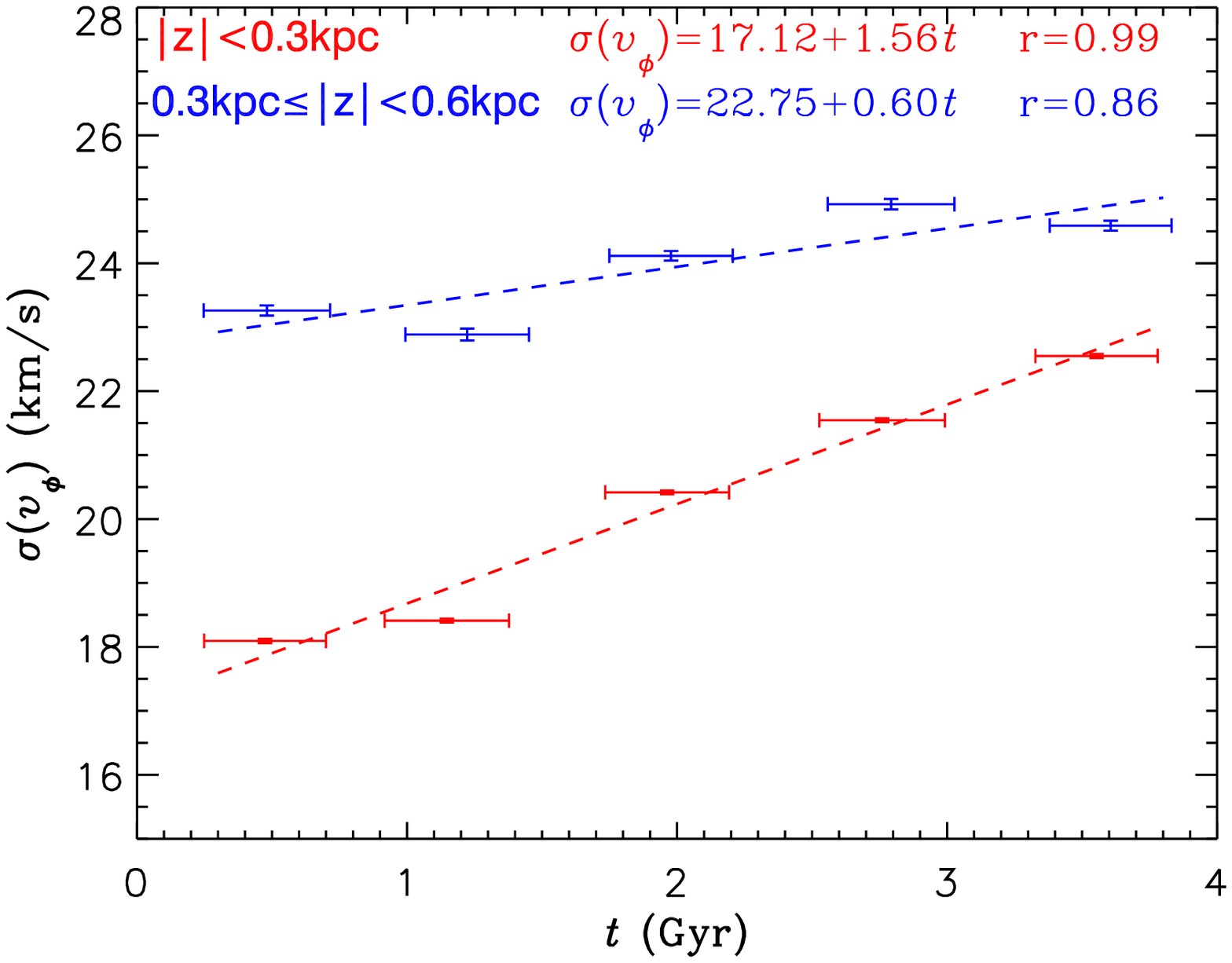}{0.5\textwidth}{(b)}
          }
\gridline{\fig{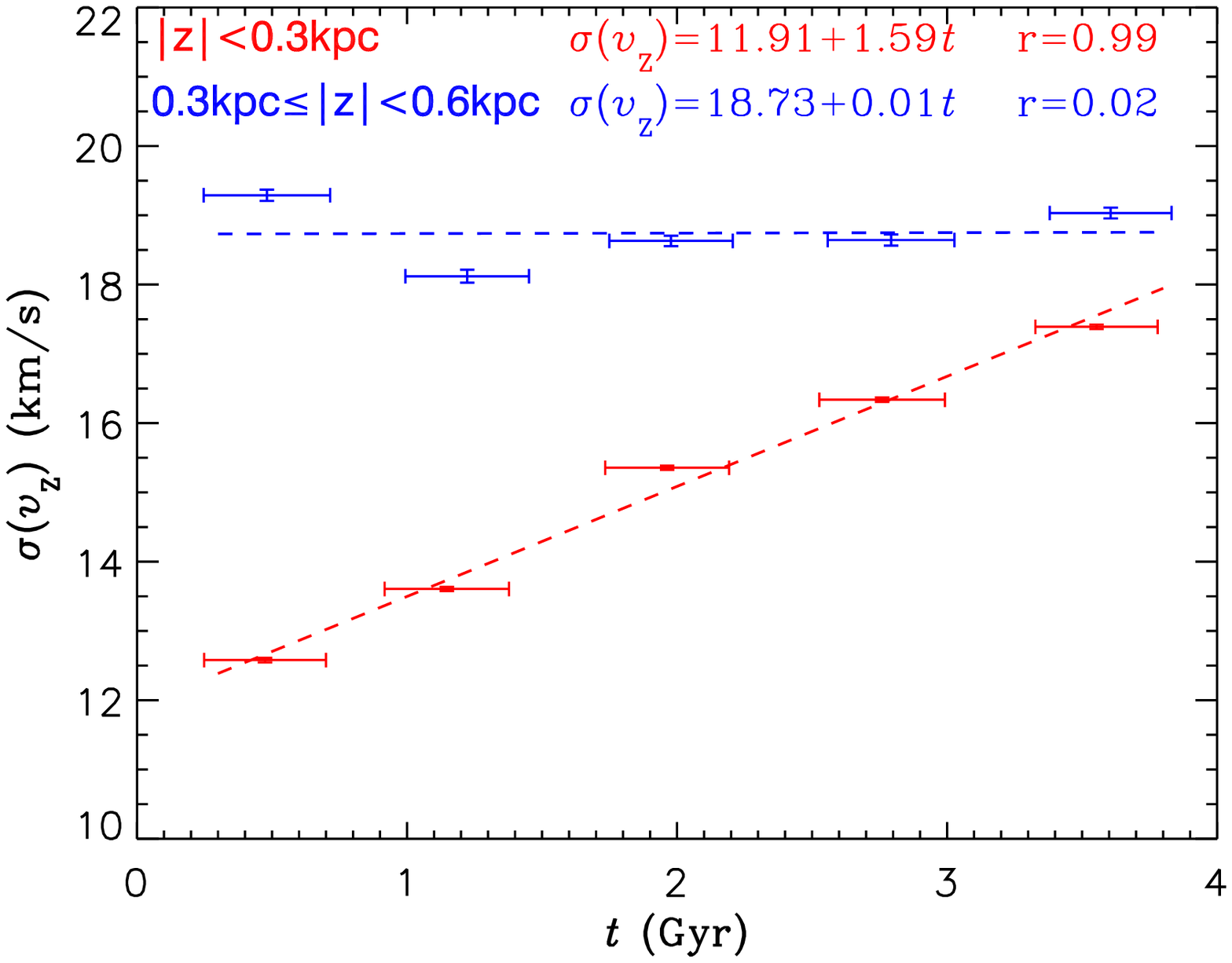}{0.5\textwidth}{(c)}
          }
\caption{Velocity dispersion versus age for two sub-samples at two different $|Z|$ bins. The red symbols represent the sub-sample with $|Z|<$ 0.3 kpc. The blue symbols represent the sub-sample with 0.3 kpc $\leq|Z|<$ 0.6 kpc. Each sub-sample was divided into five equal age bins between [0, 4.0] Gyr. The standard deviations of velocity components $\upsilon_{R}$, $\upsilon_{\phi}$ and $\upsilon_{Z}$ and the median values of the ages are shown. Horizontal error bars represent standard deviations for the ages. Vertical error bars were calculated via 100 Monte Carlo simulations. Dashed lines are linear least square fits. These relations and Pearson correlation coefficients are listed in the top of each panel.  \label{fig:AVR}}
\end{figure}

It is well known that velocity dispersion increases with age. It also correlates with orbital angular momentum $L_{Z}$, metallicity [Fe/H], and height above the plane $|Z|$ \citep{Yu2018MNRAS,Sharma2020}. Our sample has [Fe/H] $>$ -0.2 and distances from the Sun are less than 1 kpc. Thus, we only can consider two of these factors: age and height above the plane. We selected two sub-samples: $|Z|<$ 0.3 kpc and 0.3 kpc $\leq|Z|<$ 0.6 kpc. Larger $|Z|$ values could not be considered because of the small number of stars. Each sub-sample was then divided into five equal age bins between [0, 4.0] Gyr. The standard deviations of velocity components were calculated for each bin. The results are shown in Figure \ref{fig:AVR}. The red trendlines and symbols represent the sub-sample with $|Z|<$ 0.3 kpc, while the blue represents the sub-sample with 0.3 kpc $\leq|Z|<$ 0.6 kpc. Figure \ref{fig:AVR} shows velocity dispersions increase with age, as expected from previous studies \citep{Nordstrom2004,Yu2018MNRAS,Sharma2020}. Thin-disk heating is believed to be the cause \citep{Mackereth2019}.

\subsection{Spiral-shaped structures in $Z-\upsilon_{Z}$ phase space in three age bins}

\begin{figure}
\gridline{\fig{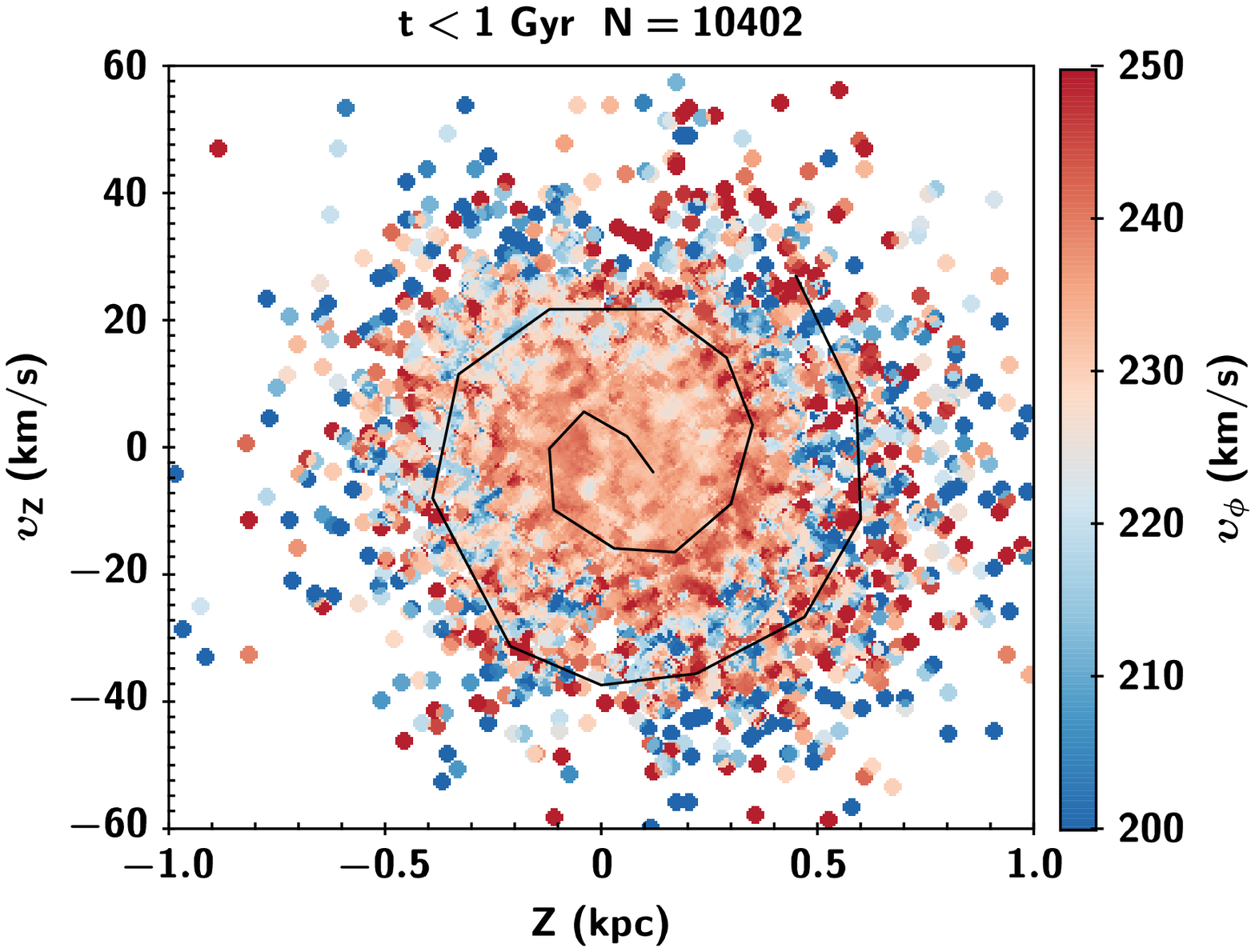}{0.4\textwidth}{(a)}
          \fig{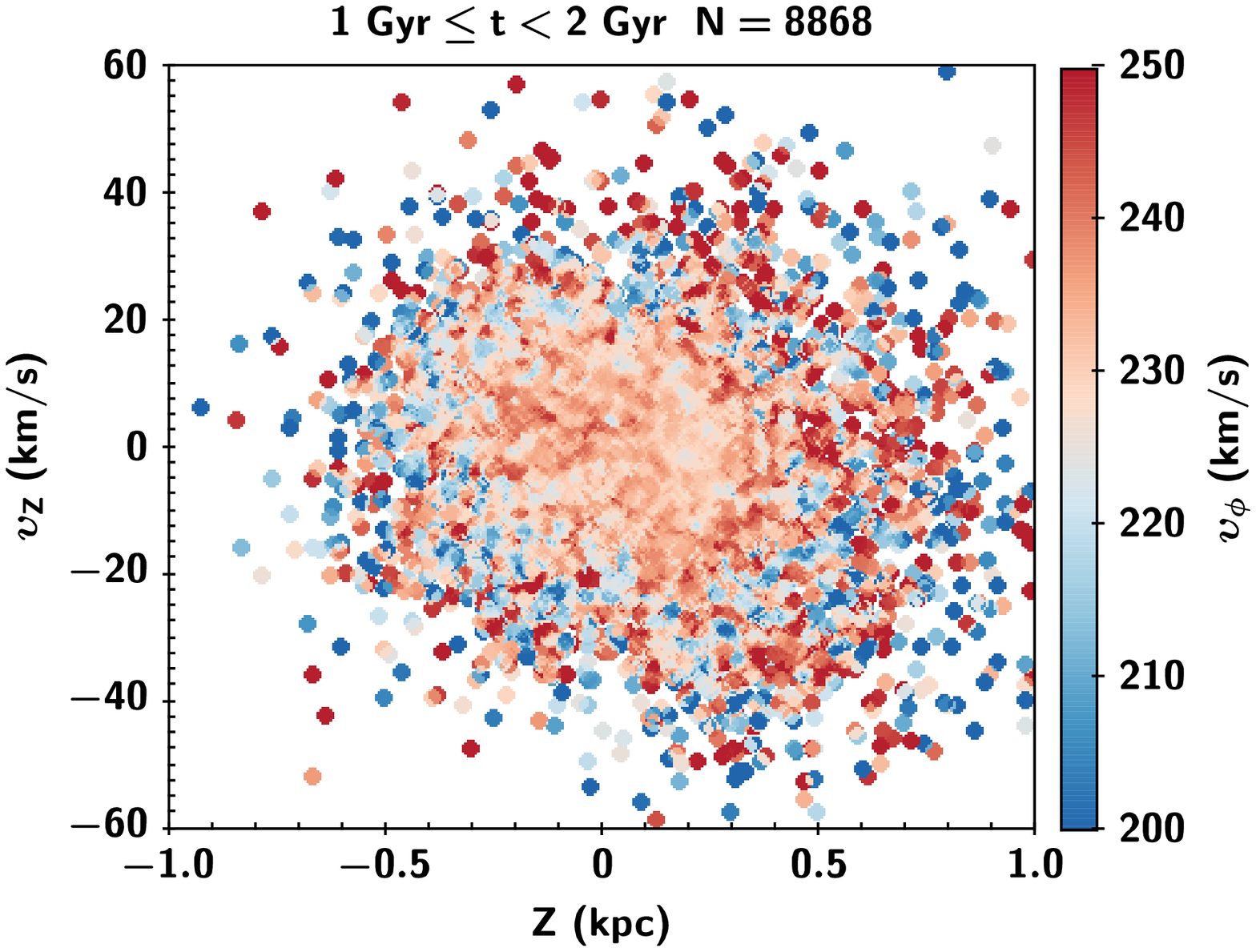}{0.4\textwidth}{(b)}
          }
\gridline{\fig{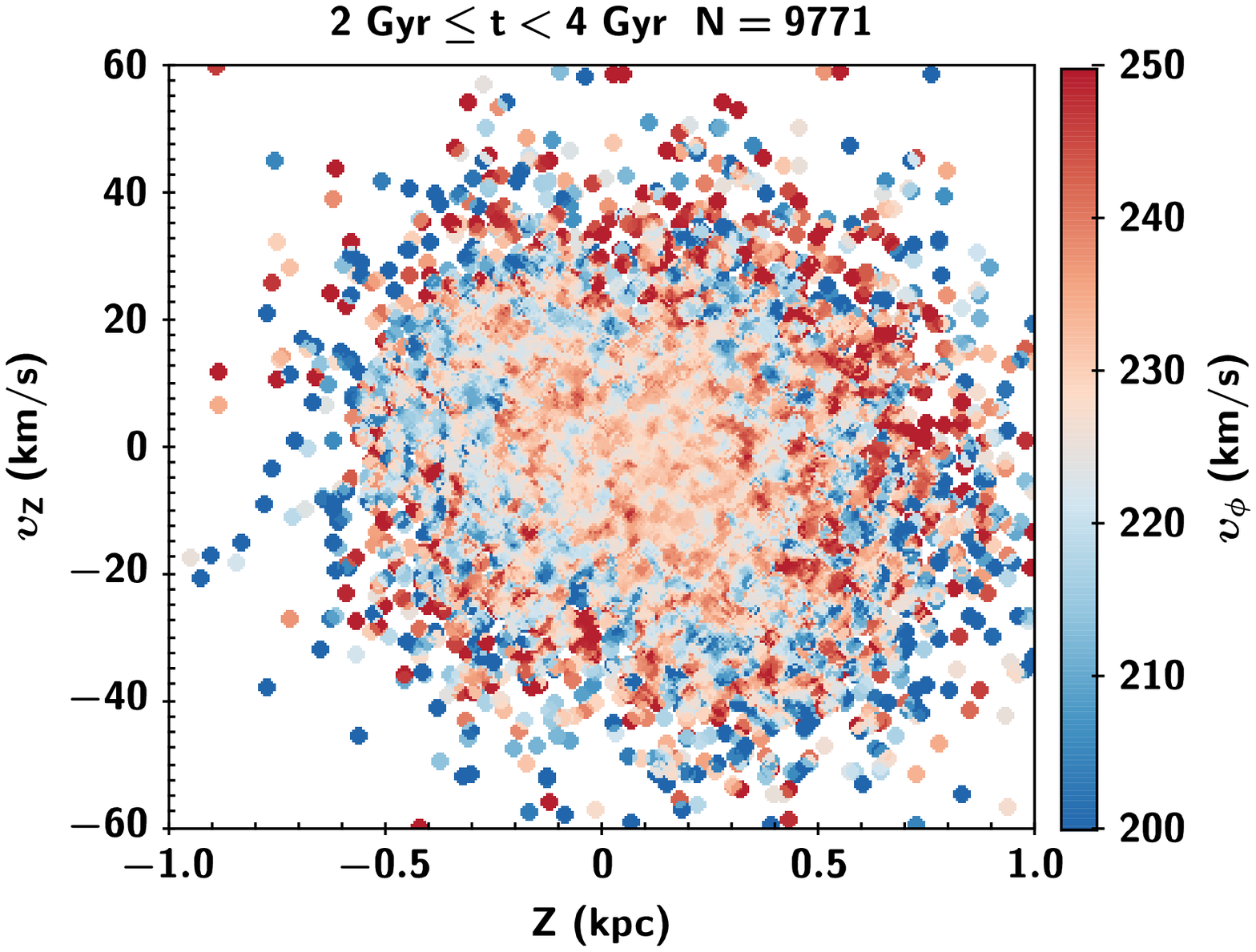}{0.4\textwidth}{(c)}
          }
\caption{Search for spiral-shaped structures in the phase space of the vertical position and velocity ($Z-\upsilon_{Z}$) for three stellar age bins: [0, 1] Gyr, [1, 2] Gyr and [2, 4] Gyr. The color represents median value of $\upsilon_{\phi}$. The hand-drawn solid line in (a) reveals where the spiral is. The age range and the number of stars at this age range are marked on the top of each panel.  \label{fig:snail diagram}}
\end{figure}

\cite{Antoja2018} discovered spiral-shaped structures in $Z-\upsilon_{Z}$ phase space, associated with vertical phase mixing of stars in the Galactic disk. It was suggested that the Galactic disk was perturbed by the Sagittarius dwarf galaxy between 300 and 900 Myr ago. In search for this pattern, we divided our sample into three stellar age bins: [0, 1] Gyr, [1, 2] Gyr and [2, 4] Gyr. Figure \ref{fig:snail diagram} presents our results in the $Z-\upsilon_{Z}$ phase space. The color scale represents median values of $\upsilon_{\phi}$. Hand-drawn solid line in Figure \ref{fig:snail diagram}(a) reveals where the spiral is. The spiral is clearly seen in the youngest age bin of [0, 1] Gyr, suggesting any vertical perturbation to the disk probably took place no later than 1.0 Gyr ago. \cite{Tian2018} concluded that the perturbation occurred no later than 0.5 Gyr ago. So our result and \cite{Tian2018} are roughly consistent. Figure \ref{fig:snail diagram} also demonstrates that as age increases, the spiral-shaped structures become less visible.

\section{Conclusion} \label{sec:conclusion}

We re-examined the CA index $\log R'\rm_{CaK}$ versus age $\log t$ relation for open clusters MS stars with 4500K $<T\rm_{eff}<$ 5000K. Quadratic functions were fitted to obtain the relation between $\log R'\rm_{CaK}$ and $\log t$. This relation was then applied to late-type MS stars with 4500K $<T\rm_{eff}<$ 5000K and [Fe/H] $>$ -0.2 to estimate their ages. Using \emph{Gaia} DR2 data, we removed stars with relative parallax errors larger than 20\% and calculated space positions ($X$, $Y$ and $Z$) and velocities ($\upsilon_{R}$, $\upsilon_{\phi}$ and $\upsilon_{Z}$). The final sample consisted of 29,041 late-type MS stars with estimated ages and kinematic information. Ages of stars in this sample range from 0.01 to 4.0 Gyr. Many references have used sub-giant stars or red giant stars to study the relation between stellar kinematics and age. Here, we have used late-type MS stars to investigate this relation. Our results complement and confirm their studies.

Our first result is the age distribution in the $R-Z$ plane. As $|Z|$ increases, the percentage of old stars increases. We also found a mild flare in stellar age distribution along the radial direction from $R$ $\sim$ 8.0 to 9.0 kpc.

Second, the age$-$velocity dispersion relation was investigated. Two sub-samples were selected: $|Z|<$ 0.3 kpc and 0.3 kpc $\leq|Z|<$ 0.6 kpc. The expected velocity dispersion increases with age was clearly detected.

Finally, the kinematics of our sample in $Z-\upsilon_{Z}$ phase space was investigated by dividing it into three stellar age bins: [0, 1] Gyr, [1, 2] Gyr and [2, 4] Gyr. A spiral-shaped structure was clearly seen in the youngest age bin. This suggests that a vertical perturbation to the disk probably took place no later than 1.0 Gyr ago. As age increases, the spiral-shaped structures were shown to fade.

\acknowledgments
This study is supported by the National Natural Science Foundation of China under grant No.11988101, 11973048, 11573035, 11625313, 11890694. This work is also supported by the Astronomical Big Data Joint Research Center, co-founded by the National Astronomical Observatories, Chinese Academy of Sciences and the Alibaba Cloud. Support from the US National Science Foundation (AST-1358787 and AST-1910396) to Embry-Riddle Aeronautical University is also acknowledged.

Guoshoujing Telescope (the Large Sky Area Multi-Object Fiber Spectroscopic Telescope LAMOST) is a National Major Scientific Project built by the Chinese Academy of Sciences. Funding for the project has been provided by the National Development and Reform Commission. LAMOST is operated and managed by the National Astronomical Observatories, Chinese Academy of Sciences.

This work uses data from the European Space Agency (ESA) mission Gaia (http://www.cosmos.esa.int/gaia), processed by the Gaia Data Processing and Analysis Consortium (DPAC, http://www.cosmos.esa.int/web/gaia/dpac/ consortium). Funding for the DPAC has been provided by national institutions, in particular the institutions participating in the Gaia Multilateral Agreement.

\end{document}